\documentclass[11pt,a4paper]{article}

\usepackage{graphicx}

\usepackage{appendix}
\usepackage{graphicx}
\usepackage[export]{adjustbox}
\usepackage{dcolumn}
\usepackage{bm}
\usepackage{epsfig}
\usepackage{epstopdf}
\usepackage{grffile}
\usepackage{color}
\usepackage{colordvi}
\usepackage{amssymb}
\usepackage{rotating}
\usepackage{lscape}
\usepackage{cite}
\usepackage{slashed}
\usepackage{dcolumn}
\usepackage{multirow}
\usepackage{float}
\usepackage{subcaption}
\usepackage[usenames]{xcolor}
\usepackage{footnote}
 \makesavenoteenv{environmentname}

\usepackage{hyperref}

\usepackage[T1]{fontenc} 
\input epsf.tex
\usepackage{amssymb}
\usepackage{amsmath}
\usepackage{amsthm}
\usepackage{amsopn}
\def\comment#1{}

\def\beq{\begin{equation}}
\def\eeq{\end{equation}}
\def\bea{\begin{eqnarray}}
\def\eea{\end{eqnarray}}

\usepackage{ulem}
\usepackage{cancel}
\usepackage{color}

\begin{document}

\title{Impact of new particles on the ratio of Electromagnetic form factors}
\author{{A.~Rafiei\footnote{e-mail:{a.rafiei@shirazu.ac.ir} }$\;,$    M.~Haghighat\footnote{e-mail:{m.haghighat@shirazu.ac.ir}} $\;$ }}
\date{\vskip5mm{ Physics Department, Shiraz University, 71946-84334, Shiraz, Iran}}

\maketitle

\begin{abstract}
We consider the electromagnetic form factors ratio in the Rosenbluth and polarization methods. We explore the impact of adding new particles as the mediators in the electron-proton scattering on these ratios. Consequently, we find some bound on the scalar coupling as $\alpha_{sc}\sim 10^{-5}$ for $m_{sc}\sim 5 MeV-2 GeV$ and $\alpha_{sc}\sim 10^{-4}-10^{-3}$ for $m_{sc}\sim 2-10 GeV$.  Meanwhile,  the vector coupling is bounded as $\alpha_v\sim 10^{-5}$ for $m_v\sim 5 MeV-1.1 GeV$ and $\alpha_v\sim 10^{-4}-10^{-3}$ for $m_v\sim 1.2-10 GeV$.  These constraints are in complete agreement with those which is found from other independent experiments.
\end{abstract}\

\section{Introduction}\label{Intro}
The electromagnetic form factors of the proton are fundamental quantities that describe the internal structure of the proton.  However, in the early 1950s, Rosenbluth was the first to express the electron-proton cross-section in terms of electric and magnetic form factors \cite{ros}.  Since then, experimental data using the Rosenbluth cross-section method has been gathered in laboratories \cite{Litt:1969my,Bartel:1973rf,Walker:1993vj,Andivahis:1994rq,Christy:2004rc,Qattan:2004ht}.  Meanwhile, in the 1970s, a new method was proposed using the concept of polarization where 
the electromagnetic form factor ratio is proportional to ratio of the transverse to longitudinal polarization of the recoiled proton~\cite{Akhiezer:1968ek,Akhiezer}. Available data~\cite{JeffersonLabHallA:1999epl,Gayou:2001qd,Punjabi:2005wq,Meziane:2010xc,Puckett:2011xg,Puckett:2017flj} based on this method had to wait for over two decades for the development of the polarization technique. However, the results have suggested that the electric form factor may be overestimated by the Rosenbluth method, leading to the proton form factor puzzle.  Many works have been made both experimentally and theoretically to understand the origin of this discrepancy. For instance, in the experimental approach one has: global reanalysis of the cross-section data \cite{Arrington:2003df}, a more precise ratio using recoil proton detection \cite{Qattan:2004ht}, and precision measurement of large wide data, which revealed the discrepancy \cite{Christy:2004rc}. However, the discrepancy still remains, indicating a significant systematic error in the theoretical calculations. Meanwhile, theoretical study is two fold: within the standard model or beyond. 

In the standard model framework, the hard two-photon exchange (TPE) hypothesis is considered a powerful explanation for the discrepancy \cite{Guichon:2003qm,Blunden:2003sp}, while soft photon radiative corrections were found to be negligible \cite{Mo:1968cg,Maximon:2000hm}. There is an assumption that hard TPE has a greater contribution in the Rosenbluth cross-section than the polarization method. However, hard TPE calculations are not model-independent, and many theoretical approaches, such as intermediate hadronic states \cite{Blunden:2005ew,Kondratyuk:2005kk,Kondratyuk:2007hc}, dispersion relations \cite{Borisyuk:2013hja,Tomalak:2014sva,Blunden:2017nby}, generalized parton distributions \cite{Afanasev:2005mp}, and phenomenological fits \cite{Chen:2007ac,Guttmann:2010au,A1:2013fsc}, are suggested to obtain hard TPE values. Such predictions suggest that hard TPE could resolve the discrepancy, while others have a different opinion \cite{Bystritskiy:2006ju,Kuraev:2007dn}. Fortunately, the hard TPE contribution can be achieved experimentally by utilizing the positron-proton to electron-proton cross-section ratio. Three experiments, OLYMPUS \cite{Henderson:2016dea}, VEPP-3 \cite{Rachek:2014fam}, and CLASS \cite{CLAS:2014xso,CLAS:2016fvy}, have attempted to measure the effects of hard TPE, but the obtained data up to $Q^2\approx 2GeV^2$ are inconsistent with theoretical predictions, though this was expected. Therefore, the hard TPE hypothesis cannot be completely ignored or confirmed, and further assessment is needed with data at higher momentum transfers, where the discrepancy is larger.

One possible explanation for the discrepancy is the existence of new physics beyond the standard model of particle physics \cite{rafiei1,rafiei2}. Many theoretical models beyond the standard model predict the existence of new particles or interactions that could affect the ratio of electromagnetic form factors.  In particular, the exchange of new particles in the electron-proton scattering could modify the scattering cross-section and affect the determination of the proton's electric and magnetic form factors. In fact, the exchange of new scalar or vector particles affect differently on the polarization that could explain the discrepancy between the two methods. 

In this paper, we aim to explore the impact of new scalar and vector particles in electron-proton scattering and their potential role in explaining the discrepancy between the Rosenbluth and polarization transfer methods.  In section 2, the theoretical framework is given very briefly.  In sections 3 and 4, we show how exchanging of new scalar and vector particles modify the electromagnetic form factor ratio in two methods. Consequently, we give some constraints on the masses and couplings of these new particles.  In section 5, summary and some concluding remarks is given.

\section{Proton Electromagnetic form factors}\label{R-P}
\begin{figure}[t]
	\centering
	\begin{tabular}{ccccc}
		\includegraphics[width=40mm,valign=c]{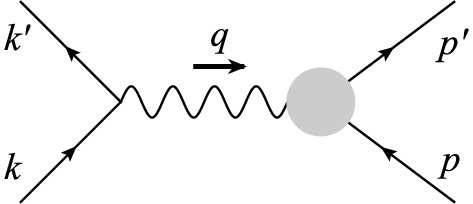}&
		+&
		\includegraphics[width=40mm,valign=c]{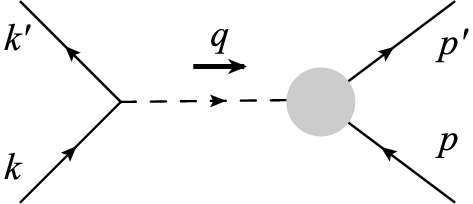}\\[6ex]
		Photon exchange && Scalar exchange \\
	\end{tabular}
	\caption{Feynman diagrams for the electron-proton elastic scattering. Bubbled vertex indicates proton target.}\label{fig.ep}
\end{figure}

In this section we introduce two experimental techniques which can be used to determine the proton form factor ratio, namely the Rosenbluth and polarization transfer methods.  To this end, we consider the
cross section of the elastic electron-proton scattering in the laboratory frame as~\cite{Greiner,Puckett}

\begin{equation}\label{cross section}
\frac{{d\sigma }}{{d\Omega }} = \frac{{\overline {\left| \cal{M}  \right|^2} }}{{64{\pi ^2}{M^2}}}{\left( {\frac{{E'_{e}}}{E_{e}}} \right)^2}\;,
\end{equation}
where $E_{e}$ and ${E'_{e}}$ are energy of the incident and scattered electrons, respectively. However, the invariant amplitude $\cal{M}$ in the one photon exchange approximation Fig.~\ref{fig.ep} can be written as 
\begin{equation}\label{photon-inv.ampl}
{{\cal{M}}_{ph}} = {e^2} J_e^\mu \frac{{{g_{\mu \nu }}}}{{{q^2}}}J_p^\nu ,
\end{equation}
where the electron current is $J_e^\mu  = \bar u(k'){\gamma ^\mu }u(k)$ and the electromagnetic current for proton which is not a point particle can be introduced as follows
\begin{equation}\label{current1}
J_\mu ^{p} = \bar u(p')[{G_M}({Q^2}){\gamma _\mu } + \frac{{{G_E}({Q^2}) - {G_M}({Q^2})}}{{2M(1 + \tau )}}({p'_\mu } + {p_\mu })]u(p)\;,
\end{equation}
where $M$ and ${Q^2} =  - {q^2}$ are the proton mass and the squared momentum transfer, respectively. Meanwhile, the electric form factor $G_E$ and the magnetic form factor $G_M$  can be defined in terms of $F_1$ (Dirac form factor) and $F_2$ (Pauli form factor) as  $G_E = F_1 - \tau F_2$ and $G_M = F_1 + F_2$ where $\tau  = \frac{{{Q^2}}}{{4{M^2}}}$ \cite{Sachs}.
 
\begin{figure}[t]
	\centerline{ \includegraphics[width=0.6\textwidth]{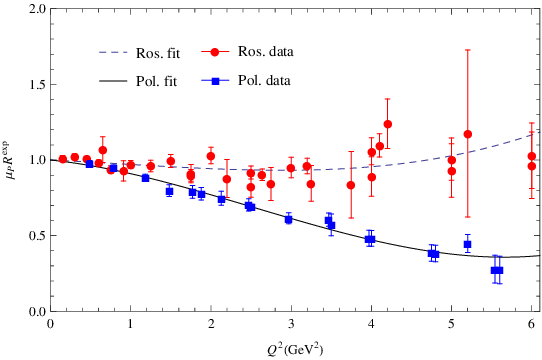}}
	\caption{Form factors ratio for Rosenbluth (red) and polarization (blue) methods.  The experimental data in Rosenbluth and polarization methods are taken from \cite{Walker:1993vj,Andivahis:1994rq,Christy:2004rc,Qattan:2004ht} and \cite{JeffersonLabHallA:1999epl,Gayou:2001qd,Punjabi:2005wq,Meziane:2010xc,Puckett:2011xg,Puckett:2017flj}, respectively. The Rosenbluth (dashed) and polarization (solid) fit are depicted from (\ref{fitros}) and (\ref{fitpol}), respectively.}
	\label{RosPol}
\end{figure}

Now the squared invariant amplitude in (\ref{cross section}), can be easily evaluated as 

\begin{equation}\label{photon-square.ampl}
\overline {|{{\cal{M}}_{ph}}{|^2}} = \frac{{{e^4}}}{{{q^4}}}{L^{\mu \nu }}{W_{\mu \nu }}  = \frac{{{e^4}}}{{{q^4}}}4{M^2}{Q^2}\left[ {\frac{{G_E^2 + \tau G_M^2}}{{1 + \tau }}{{\cot }^2}\frac{\theta }{2} + 2\tau G_M^2} \right],
\end{equation}
where $L^{\mu \nu}$ and $W_{\mu \nu }$ are the Leptonic and hadronic tensors, respectively, and $\theta$ is the scattering angle in the laboratory frame. Therefore, by using the Mott cross section for point  particles as ${\left( {\frac{{d\sigma }}{{d\Omega }}} \right)_{Mott}} = \frac{{{\alpha ^2}{{\cos }^2}\frac{\theta }{2}}}{{4{E^2}{{\sin }^4}\frac{\theta}{2}}}$ one can introduce a reduced cross section in the one-photon exchange approximation as 
\begin{equation}\label{redcross}
{\sigma _{red,ph}}=\frac{{d\sigma }}{{d\Omega }}\frac{{\varepsilon (1 + \tau )}}{{{{\left( {\frac{{d\sigma }}{{d\Omega }}} \right)}_{Mott}}}} = \varepsilon G_E^2 + \tau G_M^2,
\end{equation}
where $\varepsilon  = {\left( {1 + 2(1 + \tau ){{\tan }^2}\frac{\theta}{2}} \right)^{ - 1}}$ represents polarization of the virtual photon.  The reduced cross section (\ref{redcross}) which depends linearly on $\varepsilon$ is called the Rosenbluth cross section. Therefore, for a fixed ${Q^2}$ the $G_E^2$ can be considered as the slope and $\tau G_M^2$ is the intercept of the reduced cross section in terms of $\varepsilon$. The $G_{E}$ and $G_{M}$ can be experimentally parameterized in terms of ${Q^2}$.  For instance, the early experimental data shows a scaling parameterization dipole fit for $G_{E}$ and $G_{M}$ as
\begin{equation}
{G_E} \approx \frac{{{G_M}}}{{{\mu _p}}} \approx {\left( {1 + \frac{{{Q^2}}}{{0.71\;Ge{V^2}}}} \right)^{ - 2}},
\end{equation}
where ${\mu _p}$ is the proton magnetic moment.  Based on the data given in \cite{Walker:1993vj,Andivahis:1994rq,Christy:2004rc,Qattan:2004ht} a polynomial fit for the ratio $R = \frac{G_E}{G_M}$ can be given as
 \begin{equation}\label{fitros}
   \begin{array}{l}
    {\mu _p}{R_{Ros}} = 1 - 0.0246546\,{Q^2}\\
    \;\;\;\;\;\;\;\quad  - 0.00732527\,{Q^4} + 0.00267707\,{Q^6}
    \end{array}
 \end{equation}
In the polarization method, the ratio $R$ is determined by measuring the transverse and longitudinal  polarization (${{P_T}}$, ${{P_L}}$) of the recoiled proton in scattering of the polarized electrons from the unpolarized protons. It can be calculated by examining the antisymmetry part of $L_{\mu \nu }W^{\mu \nu }$ in the ${\hat x}$ and $\hat z$-directions as $L_{\mu \nu }^AW_A^{\mu \nu }$ after summing on the spin of the polarized electron and performing the trace as follows~\cite{Puckett}
\begin{equation}\label{pol.vec.9}
\frac{{{P_T}}}{{{P_L}}} = \frac{{L_{\mu \nu }^AW_A^{\mu \nu }(\hat x)}}{{L_{\mu \nu }^AW_A^{\mu \nu }(\hat z)}} =  - \frac{{{{G}_E}}}{{{{G}_M}}}\frac{{2M}}{Q}\cos \frac{{{\theta _B}}}{2}\;,
\end{equation}
where ${{\theta _B}}$ is the  scattering angle in the Breit frame and by substituting the scattering angle in the laboratory frame ${{\theta}}$ instead of ${{\theta _B}}$ one has~\cite{Puckett}
\begin{equation}\label{ratio}
    \frac{{{G_E}}}{{{G_M}}} =  - \frac{{{P_T}}}{{{P_L}}}\frac{{E_{e} + E'_{e}}}{{2M}}\tan \frac{\theta}{2}\;.
\end{equation}
Meanwhile, a polynomial fit for the ratio $R = \frac{G_E}{G_M}$ ~\cite{Guichon:2003qm} based on the experimental data ~\cite{JeffersonLabHallA:1999epl,Gayou:2001qd} leads to
\begin{equation}\label{fitpol}
    \begin{array}{l}
    {\mu _p}{R_{Pol}} = 1 - 0.0549647\,{Q^2}\\
    \quad \quad \quad  - 0.0407624\,{Q^4} + 0.00537502\,{Q^6}
    \end{array}
\end{equation}
As one can see in Fig.\ref{RosPol}, comparing the polynomial fits  ~(\ref{fitros}) and ~(\ref{fitpol}) show a considerable discrepancy between these methods which is increasing with the  ${Q^2}$ value.

\section{A New Scalar Particle}
In this section we consider exchanging a new scalar particle along with the ordinary electromagnetic interaction in the electron-proton scattering. The Lagrangian density for the scalar part can be considered as ~\cite{Liu:2015sba} 

\begin{equation}
	{{\cal{L}}_{sc} } =  \frac{1}{2}{(\partial \phi )^2} - \frac{1}{2}m_{sc}^2{\phi ^2} + g_{sc}^f\phi {\bar \psi _f}{\psi _f} ,
\end{equation}

 where $m_{sc}$ is the mass of the scalar particle and $g_{sc}^f$ shows the couplings with leptons and nucleons. Therefore, in addition to the electromagnetic interaction one should consider a new diagram at tree level for the electron-proton scattering as is shown in Fig. \ref{fig.ep} which leads to the total invariant amplitude as follows
\begin{equation}\label{mtot}
{{\cal{M}}_{tot}} = {{\cal{\tilde{M}}}_{ph}} + {{\cal{M}}_{sc}},
\end{equation}
where ${\cal{\tilde{M}}}_{ph}$ is defined by replacing $G_E$ by $\tilde{G}_E $ in (\ref{photon-inv.ampl}) and
\begin{equation}\label{inv.amp.s}
{{\cal{M}}_{sc}} ={g_{sc}^e}{g_{sc}^p} {J_e}\frac{{1}}{{{q^2} - m_{sc}^2}}{J_p},
\end{equation}
 where $g_{sc}^e$ and $g_{sc}^p$ show the scalar couplings with electron and proton, respectively. Furthermore, the scalar-fermion coupling leads to a current ${J_e} = \bar u(k')u(k)$ for the electron and ${J_p} = F'({q^2})\bar u(p')u(p)$ for the proton as an extended particle with the structure $F'({q^2})$.  Now one can easily show that in the vanishing electron mass limit the cross terms in the average squared total invariant amplitude is vanished and one has
\begin{equation}\label{square.total.amp}
\overline {{{\left| {{{\cal{M}}_{tot}}} \right|}^2}}  =\overline {{{\left| {{{\cal{\tilde{M}}}_{ph}}} \right|}^2}} + \overline {{{\left| {{{\cal{M}}_{sc}}} \right|}^2}},
\end{equation}
where the first term is given in (\ref{photon-square.ampl}) and 
\begin{equation}\label{square.amp.scalar}
\overline {{{\left| {{{\cal{M}}_{sc}}} \right|}^2}}  = \frac{{(g_{sc}^eg_{sc}^p)^2{{F'}^2}}}{{{{\left( {{q^2} - m_{sc}^2} \right)}^2}}}8{M^2}(k'.k)(1 + \tau ).
\end{equation}
As the cross terms in the squared of total amplitude is vanished the total cross section can be obtained by adding the scalar cross section to the photon cross section.  The reduced cross section for the electromagnetic part is given in (\ref{redcross}) while in the scalar part the corresponding term for (\ref{square.amp.scalar}) after a little algebra can be obtained as
\begin{equation}\label{cross-section-scalar}
{\sigma _{red,sc}} = \frac{1}{2}{\zeta ^2}(1 + \tau )(1 - \varepsilon ),
\end{equation}
where  $\zeta  = \frac{{{\alpha_{sc}}}}{\alpha }F'({q^2}){\left( {1 + \frac{{m_{sc}^2}}{{{Q^2}}}} \right)^{ - 1}}$ and ${\alpha_{sc}} = \frac{{{g_{sc}^e}{g_{sc}^p}}}{{4\pi }}$. Since in our model the new particles couple to the proton as a point particle therefore
the magnetic form factor at the leading order (tree level) does not receive any contribution from exchanging of such a particle. In fact, $G_E$ should be changed to $\tilde{G}_E $ and one can consider $G_M$ for the magnetic form factor without any changes in the notation with respect to the standard notation.  Therefore, one can assume the modified electromagnetic ratio as $\tilde R = \frac{{\tilde G}_E}{{G_M}}$ ($\tilde G_M=G_M$) to find
\begin{equation}\label{ph-s-cs1}
{\sigma _{tot}} = {\tilde{\sigma}_{red,ph}} + {\sigma _{red,sc}} = \varepsilon \tilde G_E^2 + \tau  G_M^2 + \frac{1}{2}{\zeta ^2}(1 + \tau )(1 - \varepsilon ),
\end{equation}

or
\begin{equation}\label{ph-s-cs2}
{\sigma _{tot}} =  G_M^2( {\varepsilon ( {{{\tilde R}^2} + \frac{1}{2}\frac{{{\zeta ^2}}}{{ G_M^2}}(1 + \tau )(\frac{{1 - \varepsilon }}{\varepsilon })} ) + \tau } ).
\end{equation}
In (\ref{ph-s-cs2}) the left hand side is experimentally compared with (\ref{redcross}) or $\sigma _{tot}=G_M^2(\varepsilon R_{Ros}^2+\tau )$
 to find the experimental value for $R_{Ros}^2$.   Therefore, the experimental Rosenbluth ratio in terms of the theoretical one, can be easily obtained as
\begin{equation}\label{ros-ratio-ph-s}
R_{Ros}^2 = {{\tilde R}^2} + \frac{1}{2}\frac{{{\zeta ^2}}}{{ G_M^2}}(1 + \tau )(\frac{{1 - \varepsilon }}{\varepsilon }).
\end{equation}
One can also find a relation between the modified electric form factor and the ordinary one as  ${G^2}_E={\tilde{G}}^2_E  + \frac{1}{2}{\zeta ^2}(1 + \tau )(\frac{{1 - \varepsilon }}{\varepsilon }) $ from (\ref{ros-ratio-ph-s}). 
The second term shows the contribution of exchanging the new scalar particle on the Rosenbluth ratio.  In contrast, the scalar particle is a spinless particle which can not have any contribution to the corresponding ratio in the polarization method~\cite{Liu:2015sba} or
\begin{equation}\label{NCpol.ratio}
{R_{Pol}} = \tilde R \;.
\end{equation}

\begin{figure}[t]
	\centering
	~ 
	\begin{subfigure}[t]{0.6\textwidth}
		\includegraphics[width=\textwidth]{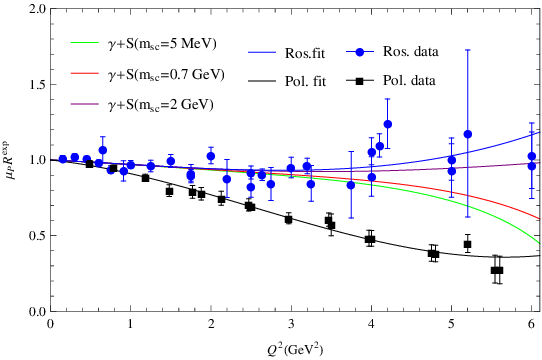}
		\caption{}
		\label{fig:second}
	\end{subfigure}
~ 
\begin{subfigure}[t]{0.6\textwidth}
	\includegraphics[width=\textwidth]{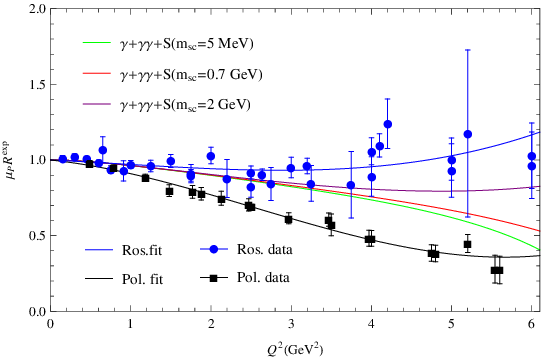}
	\caption{}
	\label{figs2ph:second}
\end{subfigure}
	\caption{The contribution of a scalar particle exchange on the ratio $R$: a) without TPE and b) with TPE, with different scalar masses.  The experimental data in Rosenbluth (blue) and polarization (black) methods are taken from \cite{Walker:1993vj,Andivahis:1994rq,Christy:2004rc,Qattan:2004ht} and \cite{JeffersonLabHallA:1999epl,Gayou:2001qd,Punjabi:2005wq,Meziane:2010xc,Puckett:2011xg,Puckett:2017flj}, respectively. The Rosenbluth (solid blue) and polarization (solid black) fit are depicted from (\ref{fitros}) and (\ref{fitpol}), respectively. }\label{s-correction-plot}
\end{figure}

The parameters  $\zeta $ and ${\alpha_{sc}}$ are free parameters of our model which can be determined by comparing the difference between (\ref{ros-ratio-ph-s}) and (\ref{NCpol.ratio})
\begin{equation}\label{ratio-diff}
R_{Ros}^2-{R_{Pol}^2} = \frac{1}{2}\frac{{{\zeta ^2}}}{{ G_M^2}}(1 + \tau )(\frac{{1 - \varepsilon }}{\varepsilon }),
\end{equation}
with the experimental discrepancy between Rosenbluth and polarization ratio.  To this end, we compare our results with the experimental data in two different cases as follows:

\textbf{First case:} In this case we assume the discrepancy is only due to the exchanging of the scalar particle.  As one can see from (\ref{ratio-diff}) the difference increases with decreasing $\varepsilon$ or larger momentum transfer. However, one can compare the obtained difference in the ratios  with the experiment to find some constraint on the free parameters of this model.  Fig \ref{fig:second} shows our results with different scalar masses in comparison with the Rosenbluth and polarization ratio along with the experimental data. For instance, by assuming $m_{sc}\sim 5 MeV$ one finds $\frac{{{\alpha_{sc}}}}{\alpha } \sim 3.4\times{10^{ -3}}$ for $\varepsilon=0.1$. These results obtained by assuming that scalar particle viewed proton as a point particle (i.e. $F'({q^2})=1$).  In fact, this form factor always appears with the coupling $\alpha_{sc}$ therefore it can be absorbed in it. 

\begin{figure}[t]
	\centering
	\begin{subfigure}[t]{0.6\textwidth}
		\includegraphics[width=\textwidth]{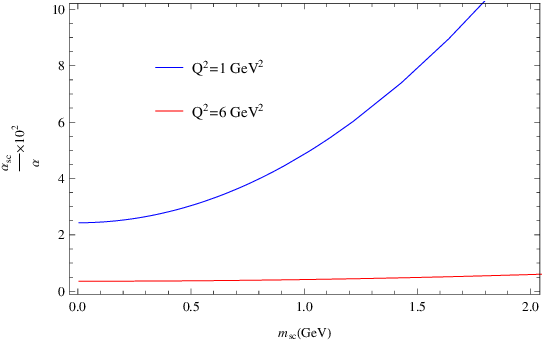}
		\caption{}
		\label{alsms1}
	\end{subfigure}
	~ 
	\begin{subfigure}[t]{0.6\textwidth}
		\includegraphics[width=\textwidth]{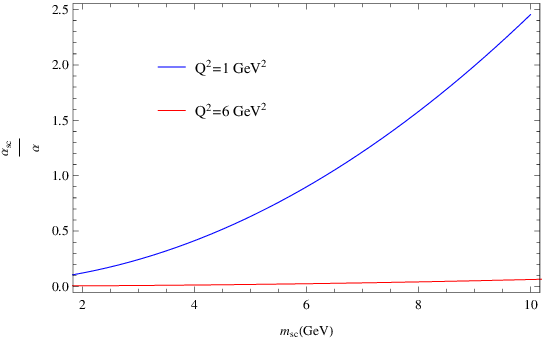}
		\caption{}
		\label{alsms2}
	\end{subfigure}
	\caption{The upper bound on the scalar coupling for different masses for $Q^2=1\, GeV^2$ (blue line) to $Q^2=6\, GeV^2$ (red line) where in:  a) $m_{sc} \approx5\, MeV-2\, GeV$ and b) $m_{sc} \approx2-10\, GeV$.  The areas above the curves in each casees are ruled out.}\label{alsms3}
\end{figure}

\textbf{Second case:} In this case as the hard TPE can partly explain the discrepency, the Rosenbluth ratio is shifted \cite{Blunden:2003sp} as 
\begin{equation}\label{R2gamma}
R_{\gamma  + \gamma \gamma }^2 = R_{Ros} ^2 - 0.014\tau \ln \left( {\frac{{{Q^2}}}{{0.65}}} \right),
\end{equation}
where $\tilde R \equiv {R_{\gamma  + \gamma \gamma }}$ shows a new ratio in the Rosenbluth method.  Therefore, adding the scalar exchange contribution (\ref{ros-ratio-ph-s}) to this case leads to a total shift in the ratio as follows  
\begin{equation}\label{shifted.ratio-2ph+s}
\tilde R^2=R_{\gamma  + \gamma \gamma  + S}^2 =R_{Ros} ^2 - 0.014\tau \ln \left( {\frac{{{Q^2}}}{{0.65}}} \right)\\
- \frac{{1}}{2}\frac{{{\zeta ^2}}}{{G_M^2}}(1 + \tau )(\frac{{1 - \varepsilon }}{\varepsilon }).
\end{equation}
  The modified Rosenbluth ratio (\ref{shifted.ratio-2ph+s}) comparing with the ordinary Rosenbluth and the polarization ratio is given in Fig. \ref{figs2ph:second} for different scalar masses.  Consequently, it can be shown that for $m_{sc} \sim 5 MeV$ the scalar coupling $\frac{{{\alpha_{sc}}}}{\alpha } \sim 2.8\times{10^{ -3}}$ can explain a large part of differences between these two methods. However, the obtained constraints in our model on the scalar coupling for different scalar masses as is shown in Fig. \ref{alsms3} is comparable with the constraints from other independent experiments which is given in Table \ref{table1}.  As is given in the table the constraints on the coupling of the scalar particle for the particle of mass around 1 MeV is about $10^{-2}$ which is already excluded in \cite{Liu:2015sba} and \cite{Tucker-Smith:2010wdq}.  In 
\cite{Liu:2015sba} the lepton-nucleon elastic scattering using all possible polarization for the incoming and outgoing particles is considered. The ratio $G_E/G_M$ in certain kinematic region is measured and a new method is presented to separately measure $G^2_E$ and $G^2_M$. Then using the constraint in \cite{Tucker-Smith:2010wdq}, which is found from $g-2$ experiment for the scalar mass $1MeV$, to show that there is not any  kinematic region to see a new scalar signal in the elastic lepton-nucleon scattering.  Nevertheless, our constraints on the scalar coupling is obtained using a different method and for a wide range of masses from a few $MeV$ to $10 GeV$.

  \begin{table}
	\fontsize{9}{9}
	\centering	\caption{The constraints on the scalar coupling from different experiments.}
	
	\begin{tabular}[b]{c c c c}
		\hline\rule[7mm]{1mm}{0mm}
		$\text{Method}$ & $m_{sc}$ & ${\alpha_{sc}}/{\alpha}$ & $\text{Reference}$\\ \hline\rule[7mm]{1mm}{0mm}
		$\textbf{Terrestrial Bounds}$ & & &\\ 
		$\text{EMFF($Q^2=1 GeV^2$)}$ & $\simeq5\,\text{MeV}-1.7\,\text{GeV}$ &$\lesssim (2-9)\times 10^{-2}$ 
		&$\text{Our Work}$\\
		$\text{EMFF($Q^2=1 GeV^2$)}$ & $\simeq 1.7-10\,\text{GeV}$ & $\lesssim 10^{-2}-2.5$ &$\text{Our Work}$\\	
	    $\text{EMFF($Q^2=6 GeV^2$)}$ & $\simeq5\,\text{MeV}-3.2\,\text{GeV}$ &$\lesssim (3-9)\times 10^{-3}$ 
		&$\text{Our Work}$\\
		$\text{EMFF($Q^2=6 GeV^2$)}$ & $\simeq 3.2-10\,\text{GeV}$ & $\lesssim (1-6)\times10^{-2}$ &$\text{Our Work}$\\
		$\text{$(g-2)_e$}$ & $\simeq 1\,\text{MeV}$ & $\lesssim 3\times 10^{-7}$ &\cite{Tucker-Smith:2010wdq}\\
		$\text{Electrophobic Scalar Boson}$ & $\simeq 10-73\,\text{MeV}$ & $\lesssim 10^{-6}-10^{-3}$ &\cite{Liu:2016qwd}\\
		$\text{Borexino-SOX}$ & $\simeq1-2\,\text{MeV}$ & $\lesssim 10^{-14}$ &\cite{Pospelov:2017kep}\\
		$\text{LSND}$ & $\simeq1-2.2\,\text{MeV}$ & $\lesssim 10^{-13}-10^{-3}$ &\cite{Pospelov:2017kep}\\
		$\text{ Underground Accelerators}$\\$\text{and Radioactive Sources}$ & $\simeq250\text{KeV}-1\text{MeV}$ & $\lesssim 10^{-10}- 10^{-7}$ &\cite{Izaguirre:2014cza}\\
		$\text{BaBar}$ & $\simeq 0.04-7\,\text{GeV}$ & $\lesssim 10^{-12}- 10^{-7}$ &\cite{BaBar:2020jma}\\
		$\text{NA64}$ & $\simeq1\,\text{MeV}-1\,\text{GeV}$ & $\lesssim10^{-10}-10^{-4}$ &\cite{NA64:2021xzo}\\
		$\text{LKB}$ & $\simeq1\,\text{MeV}-1\,\text{GeV}$ & $\lesssim10^{-9}-10^{-4}$ &\cite{Morel:2020dww}\\
		\hline\hline\rule[7mm]{1mm}{0mm}
		$ {\textbf{Cosmology Bounds}}$& &\\
		$\text{Red giant}$ & $\lesssim9\,\text{MeV}$ & $\lesssim 10^{-24}$ &\cite{Raffelt:1994ry} \\
		$\text{Solar production}$ & $ <1\,\text{MeV}$ & $\simeq 10^{-9}-10^{-13}$ &\cite{Borexino:2015axw,Pospelov:2017kep} \\
		\hline\rule[7mm]{1mm}{0mm}
	\end{tabular}\label{table1}
\end{table}
 
\section{A New Vector Particle}
In this section we examine the effect of exchanging a new vector particle along with the ordinary photon in the electron-proton scattering.  To this end, we consider a Lagrangian for the interaction of  the massive vector field with the fermions as follows~\cite{Fabbrichesi:2020wbt} 
\begin{equation}\label{v-int}
	{{\cal{L}}_v} =  - \frac{1}{4}{V^{\mu \nu }}{V_{\mu \nu }} + \frac{1}{2}m_v^2{V^\mu }{V_\mu } + g_v^f{{\bar \psi }_f}{\gamma _\mu }{\psi _f}{V^\mu },
\end{equation}
where $m_v$ is the vector mass and  ${V_{\mu \nu }} = {\partial _\mu }{V_\nu } - {\partial _\nu }{V_\mu }$ is the field strength tensor.  Meanwhile, a kinetically-mixed term \cite{Abdullahi:2023tyk} as $-\frac{g_v^f}{2eC_W}F_{\mu \nu }V^{\mu \nu }$, where $F_{\mu \nu }$ is the field strength tensor of the photon, after the diagonalization
of the gauge kinetic terms can also lead to the same interaction term given in 
(\ref{v-int}).

If we consider $g_v^e$ and $g_v^p$ for the couplings of the electron and proton with the vector particle, respectively, at the lowest order and using the current conservation (${q.J = 0}$) one finds
\begin{equation}\label{v-tree-inv.amp}
{{\cal{M}}_{v}} = g_v^eg_v^pJ_e^\mu \frac{{{g_{\mu \nu }} -\frac{{{q_\mu }{q_\nu }}}{{m_v^2}}}}{{{q^2}(1 - \frac{{m_v^2}}{{{q^2}}})}}\tilde{J}_p^\nu  = {\frac{g_v^eg_v^p}{e^2}}{(1 - \frac{{m_v^2}}{{{q^2}}})^{ - 1}}{{\cal{\tilde{M}}}_{ph}},
\end{equation}
which for the total squared invariant amplitude leads to 
\begin{equation}\label{v-a-sq-inv.amp}
\overline {|{{\cal{M}}_{tot}}{|^2}}  = {\xi ^2}(\frac{{{e^4}}}{{{q^4}}}{L^{\mu \nu }}{\tilde{W}_{\mu \nu }}),
\end{equation}
where $\xi  = 1 + \rho$,  $\rho  = \frac{{{\alpha _v}}}{\alpha }{(1 + \frac{{m_v^2}}{{{Q^2}}})^{ - 1}}$  and the coupling constant ${\alpha _v} = \frac{{g_v^eg_v^p}}{{4\pi }}$.  In fact, in this case the experimental Rosenbluth cross section can be find as 
\begin{equation}\label{extra3}
d\sigma _{_{Ros}} = {\xi ^2}d{\sigma _{theory}}.
\end{equation}
 Therefore, adding a new vector particle changes the ordinary cross section  as
\begin{equation}\label{v-cross-section}
d{\sigma _{theory}} = \frac{{d{\sigma _{Ros}}}}{{{\xi ^2}}},
\end{equation}
where in general $d{\sigma}=\varepsilon G_E^2 + \tau G_M^2=G_M^2(\varepsilon R^2 + \tau)$.  Therefore, to consider a new vector particle we define a modified ratio as $\tilde R \equiv {R_{\gamma  + V }}$ for the left hand side of (\ref{v-cross-section}) while for the right hand side of this equation $R_{Ros}^2$ shows the experimental ratio.
Since $\rho\ll 1$ by neglecting the higher orders of coupling constant one has $\xi ^{-2}\simeq 1- 2\rho$ which leads to 
\begin{equation}\label{v-shifted-ratio-0}
(\varepsilon {\tilde R^2} + \tau) = (\varepsilon R_{Ros}^2 +\tau)\left( {1 - 2\rho } \right),
\end{equation}
or for the modified Rosenbluth ratio one finds
\begin{equation}\label{v-shifted-ratio}
\tilde R^2=R_{\gamma+V}^2 = R_{Ros}^2\left( {1 - 2\rho } \right) - 2\rho \frac{\tau }{\varepsilon }.
\end{equation}
\begin{figure}[t]
	\centering
	~ 
	\begin{subfigure}[t]{0.6\textwidth}
		\includegraphics[width=\textwidth]{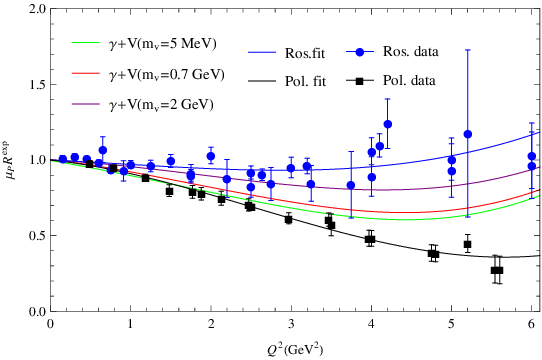}
		\caption{}
		\label{figvnon2ph:second}
	\end{subfigure}
~ 
\begin{subfigure}[t]{0.6\textwidth}
	\includegraphics[width=\textwidth]{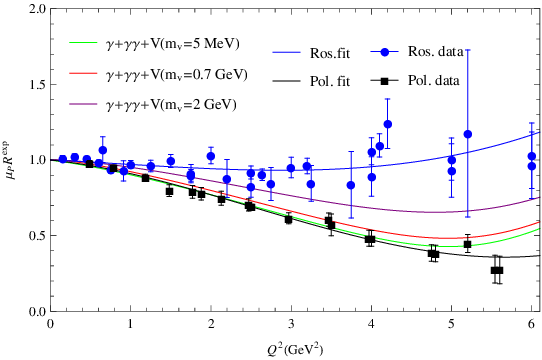}
	\caption{}
	\label{figv2ph:second}
\end{subfigure}
	\caption{The contribution of a vector particle exchange on the ratio $R$: a) without TPE and b) with TPE, with different vector masses. The experimental data in Rosenbluth (blue) and polarization (Black) methods  is taken from \cite{Walker:1993vj,Andivahis:1994rq,Christy:2004rc,Qattan:2004ht} and ~\cite{JeffersonLabHallA:1999epl,Gayou:2001qd,Punjabi:2005wq,Meziane:2010xc,Puckett:2011xg,Puckett:2017flj}, respectively. Meanwhile, Rosenbluth (solid blue) and polarization (solid black) fit are depicted using (\ref{fitros}) and (\ref{fitpol}), respectively.}\label{v-correction-plot}
\end{figure} 
In contrast, in the polarization method  as one can see from (\ref{pol.vec.9}) $\xi$ appears as a multiplicative factor in both numerator and denominator of the ratio which does not affect the polarization method in this case as follows 
\begin{equation}\label{v-shifted-ratio-pol}
AR_{Pol} = \frac{{{\xi ^2}L_{\mu \nu }^A\tilde W_A^{\mu \nu }(\hat x)}}{{{\xi ^2}L_{\mu \nu }^A\tilde W_A^{\mu \nu }(\hat z)}} = \frac{{L_{\mu \nu }^A\tilde W_A^{\mu \nu }(\hat x)}}{{L_{\mu \nu }^A\tilde W_A^{\mu \nu }(\hat z)}} = A\frac{{{{\tilde G}_E}}}{{{G_M}}} = A\tilde R,
\end{equation}
where the coefficient  $A$ can be obtained from (\ref{pol.vec.9}). Then
\begin{equation}\label{v-shifted-ratio-pol-1}
R_{Pol} = \tilde R.
\end{equation}

It is worth to mention that the normalization of the form factors are the standard one as  $G_E (0) = 1$ and $G_M = \mu_p = 2.79$ and the modified electric form factor satisfies this normalization, $\tilde{G}_E (0) =1$.

\begin{figure}
	\centerline{ \includegraphics[width=0.6\textwidth]{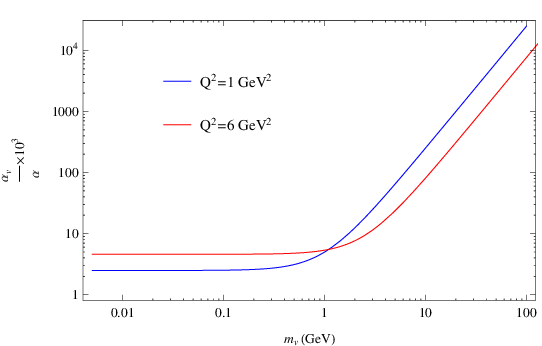}}
	\caption{The upper bound on the vector coupling for different masses is shown for $Q^2=1\, GeV^2$ (blue line) and $Q^2=6 \, GeV^2$ (red line) where areas above the curves are ruled out.}
	\label{fig.alphav-mv}
\end{figure}

Here again we can find some bounds on the free parameters of the model in two cases:
\begin{itemize}
\item 
Without the TPE, only correction from the exchange of vector particle can be considered.  In this case the effect of different masses is shown in the Fig \ref{figvnon2ph:second} where for $m_v= 5 MeV$ one obtains $\frac{{{\alpha _v}}}{\alpha } \approx 3 \times {10^{ - 3}}$. 
\item 
The TPE along with the exchange of a vector  particle, in this case the Fig. \ref{figv2ph:second} shows the effect of this new particle for different masses where with $m_v=5 MeV$ one finds $\frac{{{\alpha _v}}}{\alpha } \approx 2.5 \times {10^{ - 3}}$.
\end{itemize}
As the Table \ref{table2} shows, the obtained value $\frac{{{\alpha _v}}}{\alpha } \approx {10^{ - 3}}$ from our model is comparable with the 
constraints which is already derived in the independent experiments. Nevertheless,   $\frac{{{\alpha _v}}}{\alpha }$ in terms of  $m_v$ for different $Q^2$ is depicted in the Fig. \ref{fig.alphav-mv}.

  \begin{table}
	\fontsize{9}{9}
	\centering	\caption{The constraints on the vector coupling from different experiments.}
	
	\begin{tabular}[b]{c c c c}
		\hline\rule[7mm]{1mm}{0mm}
		$\text{Method}$ & $m_v$ &   $\alpha _v/\alpha$ & $\text{Reference}$\\ \hline\rule[7mm]{1mm}{0mm}
		$\textbf{Terrestrial Bounds}$& & \\ 
		$\text{EMFF($Q^2=1 GeV^2$)}$ & $\simeq5\,\text{MeV}-1.7\,\text{GeV}$ & $\lesssim(2-9)\times10^{-3}$ 
		&$\text{Our Work}$\\
		$\text{EMFF($Q^2=1 GeV^2$)}$ & $\simeq1.7-10\,\text{GeV}$ & $\lesssim 10^{-2}-3\times10^{-1}$ 
		&$\text{Our Work}$\\
		$\text{EMFF($Q^2=6 GeV^2$)}$ & $\simeq5\,\text{MeV}-2.6\,\text{GeV}$ & $\lesssim(4-9)\times10^{-3}$ 
        &$\text{Our Work}$\\
        $\text{EMFF($Q^2=6 GeV^2$)}$ & $\simeq2.6-10\,\text{GeV}$ & $\lesssim (1-8)\times10^{-2}$ 
        &$\text{Our Work}$\\	
		$\text{$(g-2)_e$}$ & $\simeq1\text{MeV}$ & $\lesssim6.4\times 10^{-7}$ &\cite{Tucker-Smith:2010wdq}\\
		$\text{$(g-2)_e$}$ & $\simeq1\,\text{MeV}-1\,\text{GeV}$ & $\lesssim 10^{-8}-10^{-3}$ &\cite{Abdullahi:2023tyk}\\
		$\text{LHCb}$ & $\simeq214-740\,\text{MeV}$ & $\lesssim10^{-8}-10^{-7}$ &\cite{LHCb:2019vmc}\\
		$\text{LHCb}$ & $\simeq10.6-30\,\text{GeV}$ & $\lesssim 10^{-6}-10^{-5}$ &\cite{LHCb:2019vmc}\\
		$\text{Belle II}$ &$\simeq$ $1\,\text{MeV}-5\,\text{GeV}$ & $\lesssim10^{-5}-10^{-2}$ &\cite{Belle-II:2022yaw,Abdullahi:2023tyk}\\
		$\text{KLOE-2}$ & $\simeq30-420\,\text{MeV}$ & $\lesssim10^{-5}$ &\cite{KLOE-2:2011hhj,KLOE-2:2012lii}\\
		$\text{KLOE-2}$ & $\simeq520-985\,\text{MeV}$ & $\lesssim10^{-5}-10^{-7}$ &\cite{KLOE-2:2014qxg,KLOE-2:2016ydq}\\
		$\text{BaBar}$ & $\lesssim3\,\text{GeV}$ & $\lesssim10^{-6}$ &\cite{BaBar:2017tiz}\\
		$\text{A1}$ & $\simeq40-300\,\text{MeV}$ & $\lesssim10^{-6}$ &\cite{Merkel:2014avp}\\
		$\text{DIS}$ & $\lesssim10\,\text{GeV}$ & $\lesssim10^{-3}$ &\cite{Thomas:2021lub}\\
		$\text{EWPO}$ & $\lesssim10\,\text{GeV}$ & $\lesssim10^{-3}$ &\cite{Curtin:2014cca}\\
		$\text{NA64}$ & $\simeq1-390\,\text{MeV}$ & $\lesssim10^{-9}-10^{-4}$ &\cite{NA64:2021acr}\\
		$\text{MUonE}$ & $\simeq$ $5\,\text{MeV}-1\,\text{GeV}$ & $\lesssim10^{-10}-10^{-6}$ &\cite{GrillidiCortona:2022kbq}\\
		\hline\hline\rule[7mm]{1mm}{0mm}
		$ {\textbf{Cosmology Bounds}}$& &\\
		\,
		$\text{SN1987A}$ & $\simeq 15-120\,\text{MeV}$ & $\lesssim10^{-18}$ &\cite{Chang:2016ntp}\\
		$\text{BBN,CMB}$ & $\simeq1\,\text{MeV}-10\,\text{GeV}$ & $\simeq10^{-36}-10^{-22}$ &\cite{Fradette:2014sza} \\
		\hline\rule[7mm]{1mm}{0mm}
	\end{tabular}\label{table2}
\end{table}

\section{Conclusion}\label{Con}
We investigated the impact of new particle exchange on the ratio of electromagnetic form factors in electron-proton scattering and its role in resolving the discrepancy between the Rosenbluth and polarization transfer methods. Our analysis focused on two scenarios, the first being the exchange of a spin-less particle, and the second being the exchange of a vector particle. In the case of a scalar mediator, we found that only the Rosenbluth method can be altered to determine the form factor ratio using (\ref{ros-ratio-ph-s}), while constraints on the mass and coupling of the scalar particle were obtained in two scenarios, one where the particle is demanded to explain the discrepancy and the other where it partly explains the discrepancy after accounting for TPE effects, see Fig. (\ref{s-correction-plot}). We presented our results for various scalar masses and compared them with the Rosenbluth and polarization transfer methods, and the constraints on the scalar particle's coupling were shown in Fig. (\ref{alsms3}). In addition, we explored the effects of a vector mediator on the form factor ratio, and the results were compared with those obtained from the Rosenbluth and polarization transfer methods, as shown in Fig. (\ref{v-correction-plot}). The constraints on the vector coupling were also depicted  in terms of the vector particle mass for different $Q^2$, as illustrated in Fig. (\ref{fig.alphav-mv}). We compared the obtained constraints with those from other experiments, including the electron g-2, the LHCb, Belle-II, DIS, EWPO, the LSND, ... and found that our results were consistent with these constraints, see Tables \ref{table1} and \ref{table2}.
In fact, for the scalar coupling the mass range as is presented in Table \ref{table1} are mostly in the MeV range in comparison with our limits in the range 5 MeV to 10 GeV. Meanwhile, for the vector coupling the mass range in different experiments as is presented in Table \ref{table2} is extended to 30 GeV but this mass range has not a complete overlap with ours i.e. 5 MeV to 10 GeV.  For example, in the LHCb \cite{LHCb:2019vmc} which gives one of the most stringent limits, these range are 214 MeV to 740 MeV and 10.6 GeV to 30 GeV.  Overall, our obtained limit is in agreement with the DIS \cite{Thomas:2021lub} and EWPO \cite{Curtin:2014cca} that provides valuable insights into the impact of new particle exchange on electron-proton scattering and its potential role in resolving the discrepancies between the different methods used to determine the electromagnetic form factor ratio.


\end{document}